\documentclass[acus]{JAC2000} 
            
\usepackage{graphicx}

\begin{document} 
\title{\flushright{PSN T02}\\[15pt] \centering EXPERIMENTAL STATUS
REPORT ON VECTOR MESON 
SPECTROSCOPY
\thanks{Invited talk at the workshop ``$e^+e^-$ Physics at Intermediate 
Energies'', SLAC, Stanford, April,30 - May,2, 2001. 
Author is grateful to SLAC for the support.}} 
 
\author{V.Ivanchenko, BINP, Novosibirsk, Russia}

\maketitle 
 
\begin{abstract} 
The experimental status of light vector meson spectroscopy
is discussed. The last results of $e^+e^-$ experiments 
obtained at the VEPP-2M collider in 
Novosibirsk are described and the comparison with the
old data in the mass
region from $1~GeV$ to $2.5~GeV$ is performed.
\end{abstract} 
 
\section{INTRODUCTION}  

For the first time 
the $e^+e^-$  spectroscopy study was performed in Novosibirsk in 60th at
the VEPP-2 collider. The shape of the $\rho(770)$ resonance 
have been measured \cite{Sidorov}. Since a lot of different experiments
for spectroscopy 
have been done \cite{PDG00} and as a rule the
most precise data on vector meson parameters were obtained
in $e^+e^-$ production. The current status of the vector meson 
spectroscopy is following:
\begin{itemize}
\item  
All main states of $q\overline{q}$ systems are established.
\item  
Charmonium and bottomonium families are well known.
\item  
Excitation states of $q\overline{q}$ system for $u, d, s$
quarks are not well established.
\item  
There are evidences for existence of $K\overline{K}$ or
4-quarks states in the vector meson
decays \cite{SNDpi0pi0g0,CMDpi0pi0g,SNDpi0etag,SNDpi0pi0g}.
\item  
There are evidences for existence of $N\overline{N}$ or
6-quarks states \cite{Zallo}.
\end{itemize}
The main problems of the light vector meson spectroscopy 
connect with the fact that in the mass region 
$2E=1.4\div 2.5~GeV$
total integrated luminosity $\simeq 2~pb^{-1}$ was collected at DCI
and ADONE. This statistic is incompatible with that
collected in the energy regions of the
charmonium and bottomonium families.

At the contrary, in the low energy region from the hadron production
threshold to $1.4~GeV$ the systematic studies have been performed 
in Novosibirsk at the $e^+e^-$ collider VEPP-2M.
It was in operation from 1974 to 2000 and 
the total integrated luminosity $\simeq 80~pb^{-1}$ was collected.
Important measurements were done by OLYA \cite{OLYAK+K-,OLYApi2}
and ND \cite{ND} experiments, but
the main part of integrated luminosity were taken by the 
CMD-2 \cite{CMD2} and SND \cite{SND} experiments.
Now the experimental 
program is finished, and the final data analysis is in a progress.

\section{PRODUCTION OF LIGHT VECTOR MESONS IN ELECTRON-POSITRON
COLLISIONS}  

Main advantages of the experiments on
vector mesons production in $e^+e^-$ annihilation
are following: clean initial state with the well defined quantum numbers, 
high mass resolution, good conditions for an exclusive reactions
study. The main problem of the $e^+e^-$ data analysis connect with
uncertainties in the 
interference between several resonances that often introduces
model dependence into final results (for example \cite{SNDphi}). There are
also model dependences of the data analysis 
\cite{GS,HLS,fresh,BEN1,Achasov}, which can be resolved only
after significant increasing of experimental statistic.

\subsection{$e^+e^-\to\pi^+\pi^-$ cross section}

\begin{figure}
\includegraphics[width=80mm]{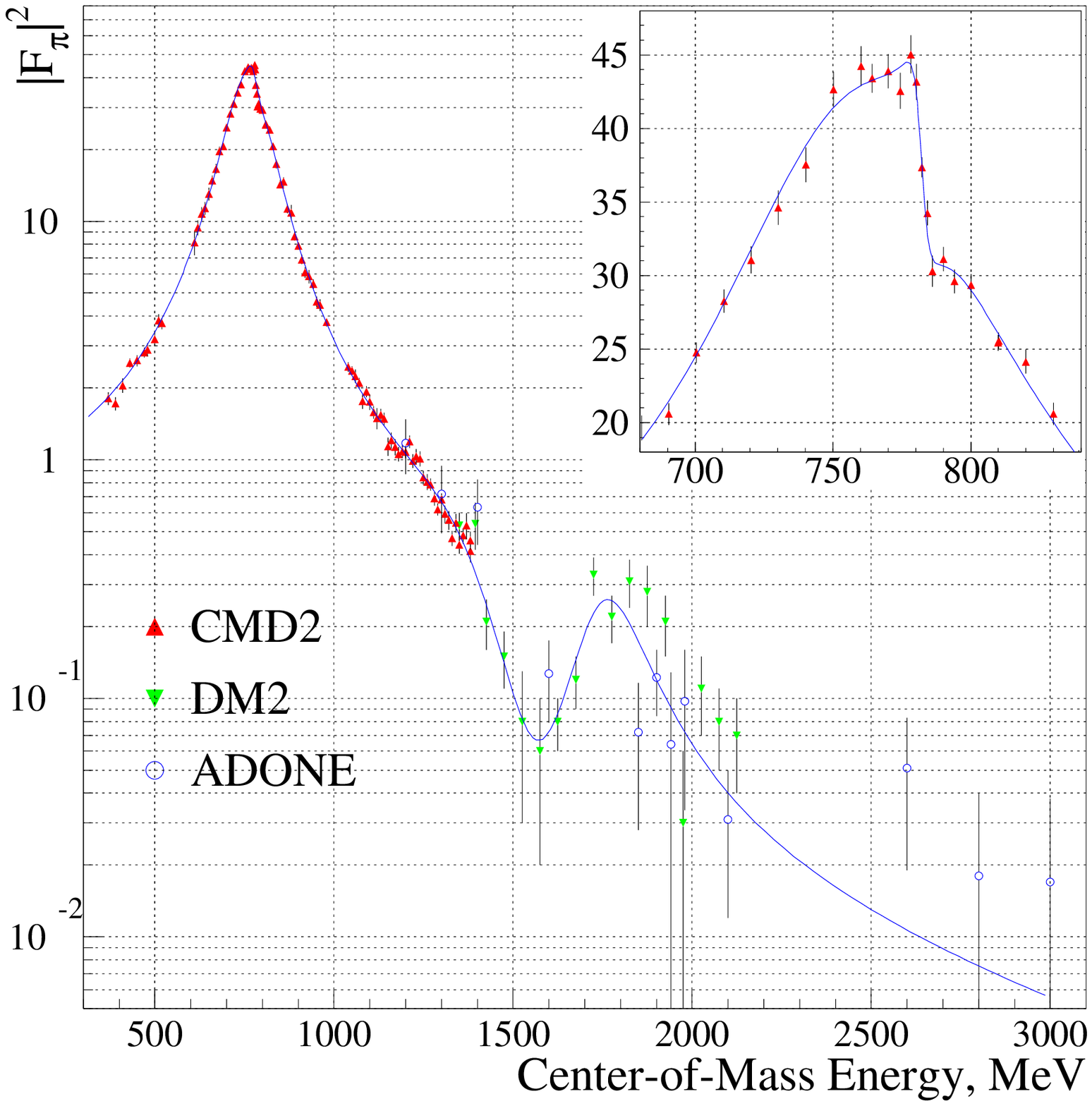} 
\caption
{The $e^+e^-\to\pi^+\pi^-$ cross section.} 
\label{pi2} 
\end{figure}

The precise  measurement of the two pion production 
cross section have been performed for many
years \cite{PDG00,OLYApi2,CMDpi2}.
The systematic uncertainty of 0.6~\% is achieved in the 
last CMD-2 experiment \cite{Logoshenko} in the energy range below $1~GeV$.
For higher energies the results are not so precise, but
DM2 data \cite{DM2pi2} strongly emphasise the signal of
$\rho(1700)$ (Fig.\ref{pi2}). There is some wide enhancement in the the cross
section around $1.25~GeV$ which may be taken as an evidence for
the $\rho(1250)$ resonance,
but at the same time another models are 
discussed \cite{HLS,BEN2}.  

\subsection{$e^+e^-\to\pi^+\pi^-\pi^0$ cross section}

\begin{figure}
\includegraphics[width=75mm]{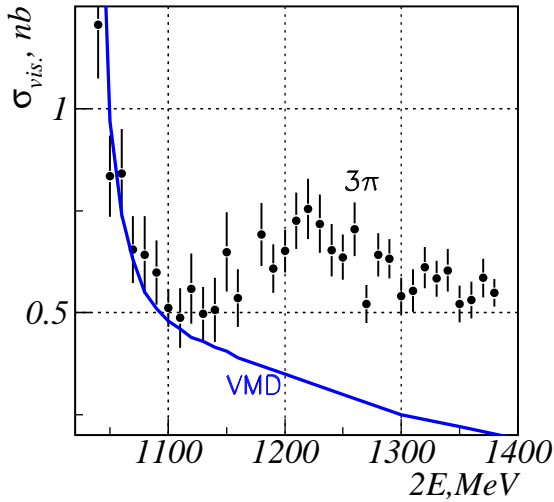} 
\caption
{Visible cross section of the reaction $e^+e^-\to\pi^+\pi^-\pi^0$.} 
\label{pi3v} 
\end{figure}

\begin{figure}
\includegraphics[width=75mm]{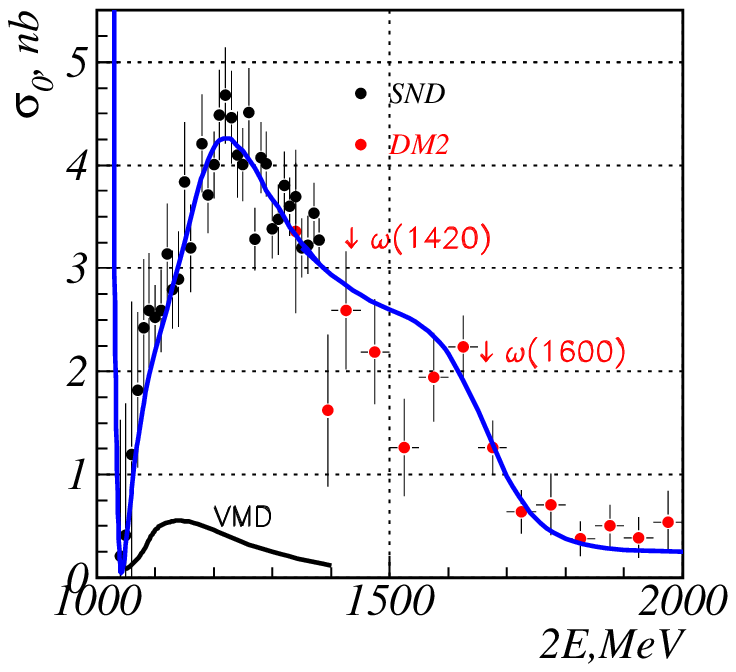} 
\caption
{The $e^+e^-\to\pi^+\pi^-\pi^0$ cross section.} 
\label{pi3} 
\end{figure}

The main contributions to
three pion production cross section at low energy come from the
$\omega(782)$ and $\phi(1020)$ resonances. It is well known that the 
interference between these resonances are destructive \cite{SNDphi}.
For many years in the energy region above $1~GeV$ the experimental 
data was not so precise \cite{ND,DM2pi3}. The last SND
measurement \cite{SNDpi3} shows that there are visible
peak in the cross section at $1.25~GeV$ (Fig.\ref{pi3v}).   
After applying the radiative corrections and 
the detection efficiencies the total cross section was obtained 
 in which the clear resonance signal is seen. 
Taking into account the data below $1~GeV$ and 
the DM2 data \cite{DM2pi3} the set of fits were performed \cite{SNDpi3}.
The best fit  (Fig.\ref{pi3})
requires contributions of $\omega$, $\phi$, $\omega(1200)$,
and $\omega(1650)$ with the relative phases (+), (-), (-), (+). 


\subsection{$e^+e^-\to\pi^+\pi^-\pi^+\pi^-$ cross section}

\begin{figure}
\includegraphics[width=80mm]{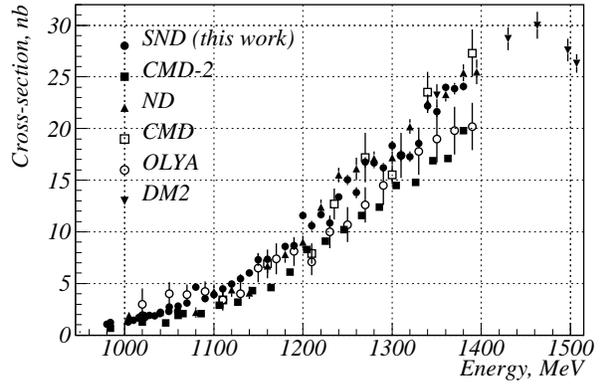} 
\caption
{The $e^+e^-\to\pi^+\pi^-\pi^+\pi^-$ cross section with the
recent VEPP-2M data \cite{SNDpi4}.} 
\label{pi4c} 
\end{figure}

The four charged pion production was studied by many groups \cite{PDG00}.
The most detailed investigation have been reported by CMD-2 
\cite{CMDa1pi}. In this work the PWA analysis have been performed
and it was shown that the  $a_1(1260)\pi$
intermediate state dominates in the energy region
below $1.4~GeV$. The SND results \cite{SNDpi4} 
confirm the CMD-2 data (Fig.\ref{pi4c}). 

\subsection{$e^+e^-\to\pi^+\pi^-\pi^0\pi^0$ cross section} 

\begin{figure}
\includegraphics[width=80mm]{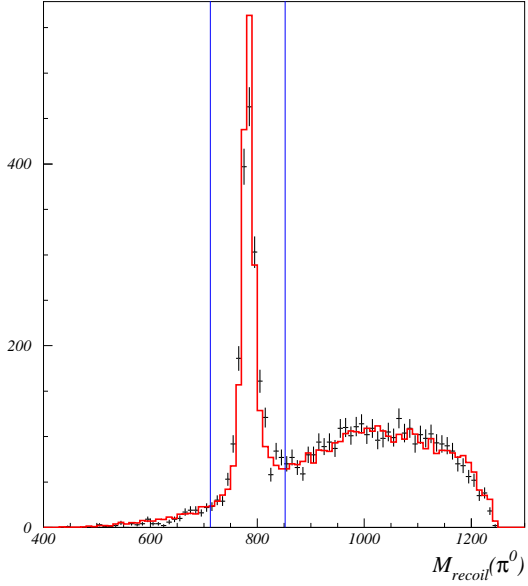} 
\caption
{The invariant mass of $\pi^+\pi^-\pi^0$ in the reaction 
$e^+e^-\to\pi^+\pi^-\pi^0\pi^0$ \cite{CMDa1pi}.} 
\label{pi4o} 
\end{figure}

\begin{figure}
\includegraphics[width=80mm]{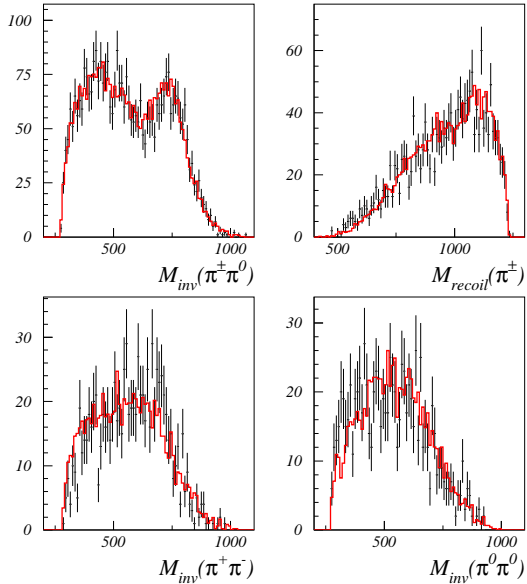} 
\caption
{The two pions invariant mass in the reaction 
$e^+e^-\to\pi^+\pi^-\pi^0\pi^0$ \cite{CMDa1pi}.} 
\label{pi4a1} 
\end{figure}

\begin{figure}
\includegraphics[width=80mm]{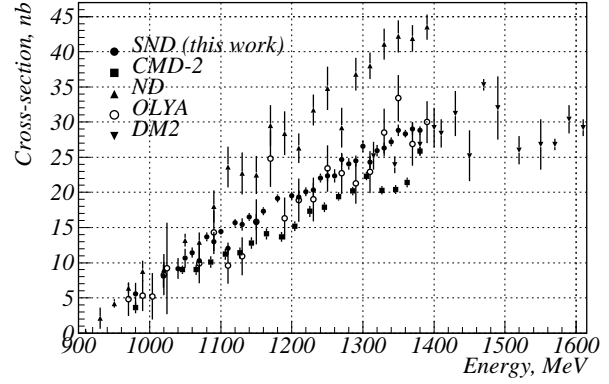} 
\caption
{The $e^+e^-\to\pi^+\pi^-\pi^0\pi^0$ cross section with the
recent VEPP-2M data \cite{SNDpi4}.} 
\label{pi4n} 
\end{figure}

Using the PWA analysis of the reaction
$e^+e^-\to\pi^+\pi^-\pi^0\pi^0$
CMD-2 obtained \cite{CMDa1pi} 
that  $a_1(1260)\pi$ and $\omega\pi^0$ 
intermediate states dominate in the  
reactions mechanism (Figs.\ref{pi4o},\ref{pi4a1}).
The recent SND data \cite{SNDpi4} are in  agreement
with the CMD-2 results within  the systematic uncertainty of the experiments
(Fig.\ref{pi4n}). 

\subsection{$e^+e^-\to\pi^+\pi^-\pi^+\pi^-\pi^0$ cross section} 

\begin{figure}
\includegraphics[width=70mm]{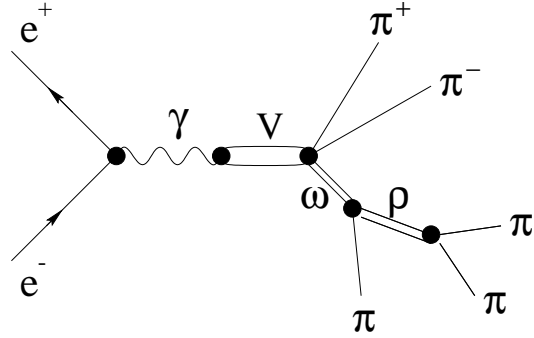} 
\caption
{The $e^+e^-\to\omega\pi^+\pi^-$ main diagram.} 
\label{d_om} 
\end{figure}

\begin{figure}
\includegraphics[width=70mm]{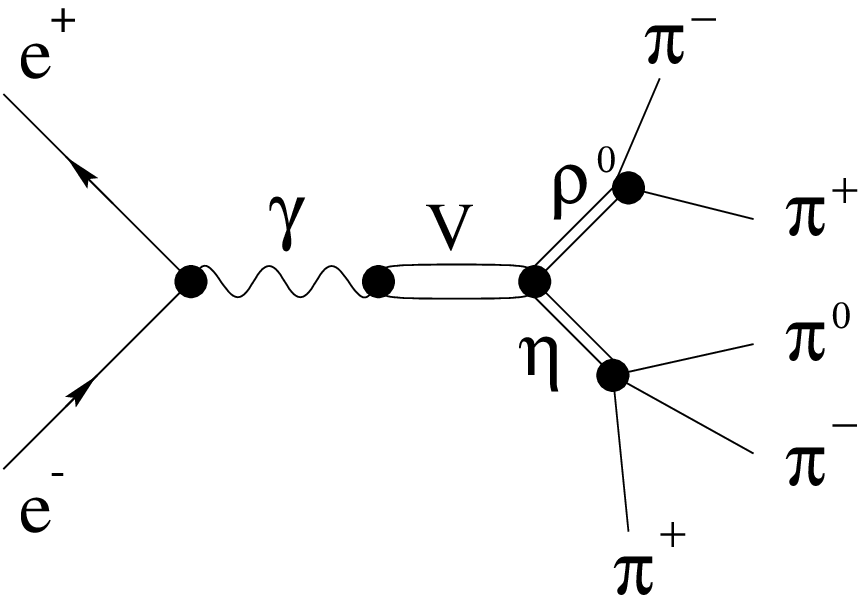} 
\caption
{The $e^+e^-\to\eta\pi^+\pi^-$ main diagram.} 
\label{d_eta} 
\end{figure}

The five pions production cross section have been studied
by CMD-2 \cite{CMDpi5} and DM2 \cite{DM2pi3}. 
It was shown that tree diagrams 
(Figs.\ref{d_om},\ref{d_eta}) dominate in these reactions.
In the $\omega\pi^+\pi^-$ cross section the clear peak of
the $\omega(1650)$ is seen and probably some contribution
of the $\omega(1200)$ exists.
In the $\eta\pi^+\pi^-$ reaction the clear peak of
$\rho(1450)$ determines the cross section shape
but some contribution of $\rho(1700)$ is not excluded.

\begin{figure}
\includegraphics[width=80mm]{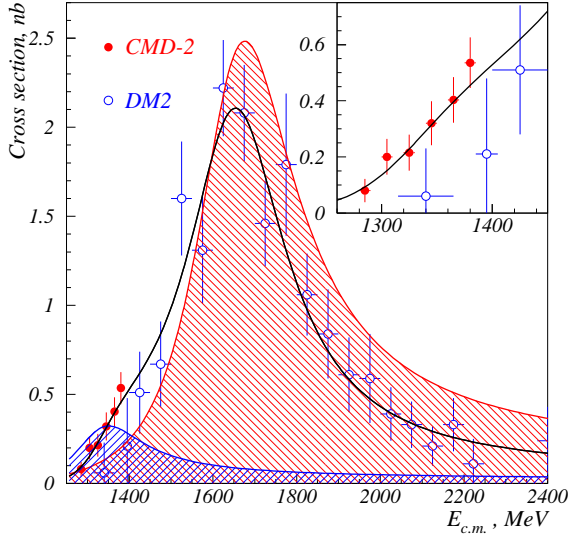} 
\caption
{The $e^+e^-\to\omega\pi^+\pi^-$ cross section.} 
\label{ompipi} 
\end{figure} 

\begin{figure}
\includegraphics[width=80mm]{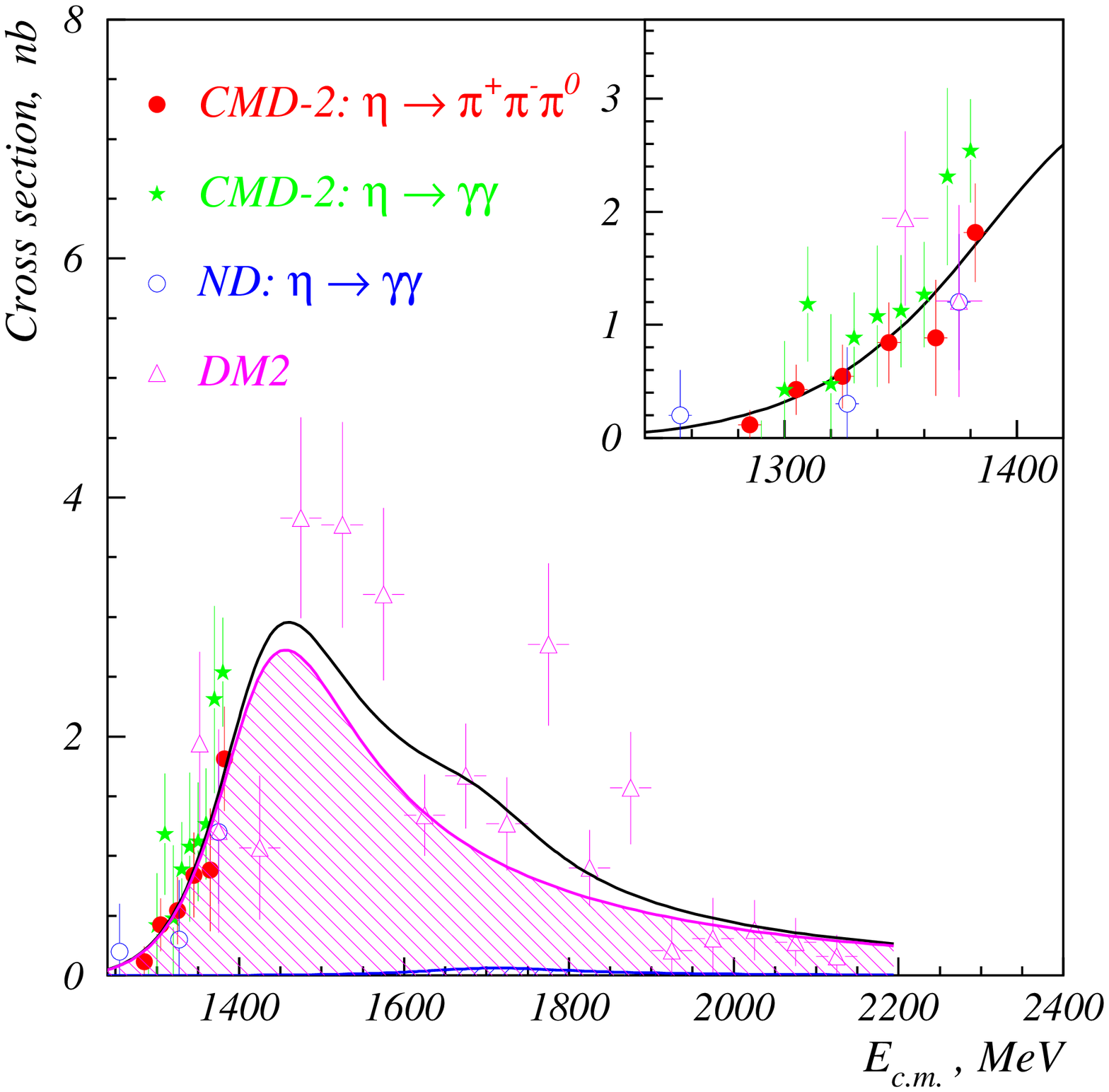} 
\caption
{The $e^+e^-\to\eta\pi^+\pi^-$ cross section.} 
\label{etapipi} 
\end{figure}

\subsection{$e^+e^-\to\omega\pi^0$ cross section}

\begin{figure}
\includegraphics[width=70mm]{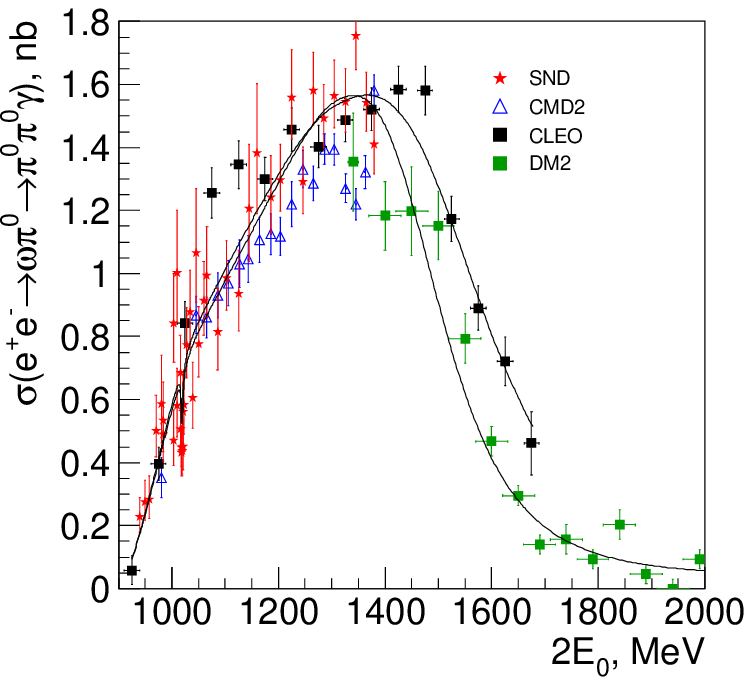} 
\caption
{The $e^+e^-\to\omega\pi^0$ cross section.} 
\label{omegapi0} 
\end{figure}

The main reaction channel $e^+e^-\to\omega\pi^0\to\pi^+\pi^-\pi^0\pi^0$
is seen in the four pion final state but less systematic
uncertainty in the cross section measurement was achieved
by SND using the $e^+e^-\to\omega\pi^0\to\pi^0\pi^0\gamma$
reaction \cite{SNDomegapi0}. Combining SND data with the
data of DM2 \cite{DM2omegapi0} and CLEOII \cite{CLEOomegapi0}
the fit of the cross section was performed (Fig.\ref{omegapi0}).
Note, that there is a systematic bias between the DM2 and CLEOII 
data, which can be connected with a normalisation problem or
with the bias in the energy scale.

\subsection{$e^+e^-\to\eta\gamma$ cross section} 

First indication of a radiative decay of radial 
excitations of light vector mesons was found out by 
CMD-2 \cite{CMDetag}. Two events 
of the reaction $e^+e^-\to\eta\gamma$ were identified.
The estimated production cross section is in agreement
with the data of CMD-2 \cite{CMDpi5} and DM2 \cite{DM2etapipi}
for the reaction $e^+e^-\to\eta\pi^+\pi^-$.

\subsection{$e^+e^-\to K_S K_L, K^+K^-$ cross sections}

\begin{figure} 
\includegraphics[width=80mm]{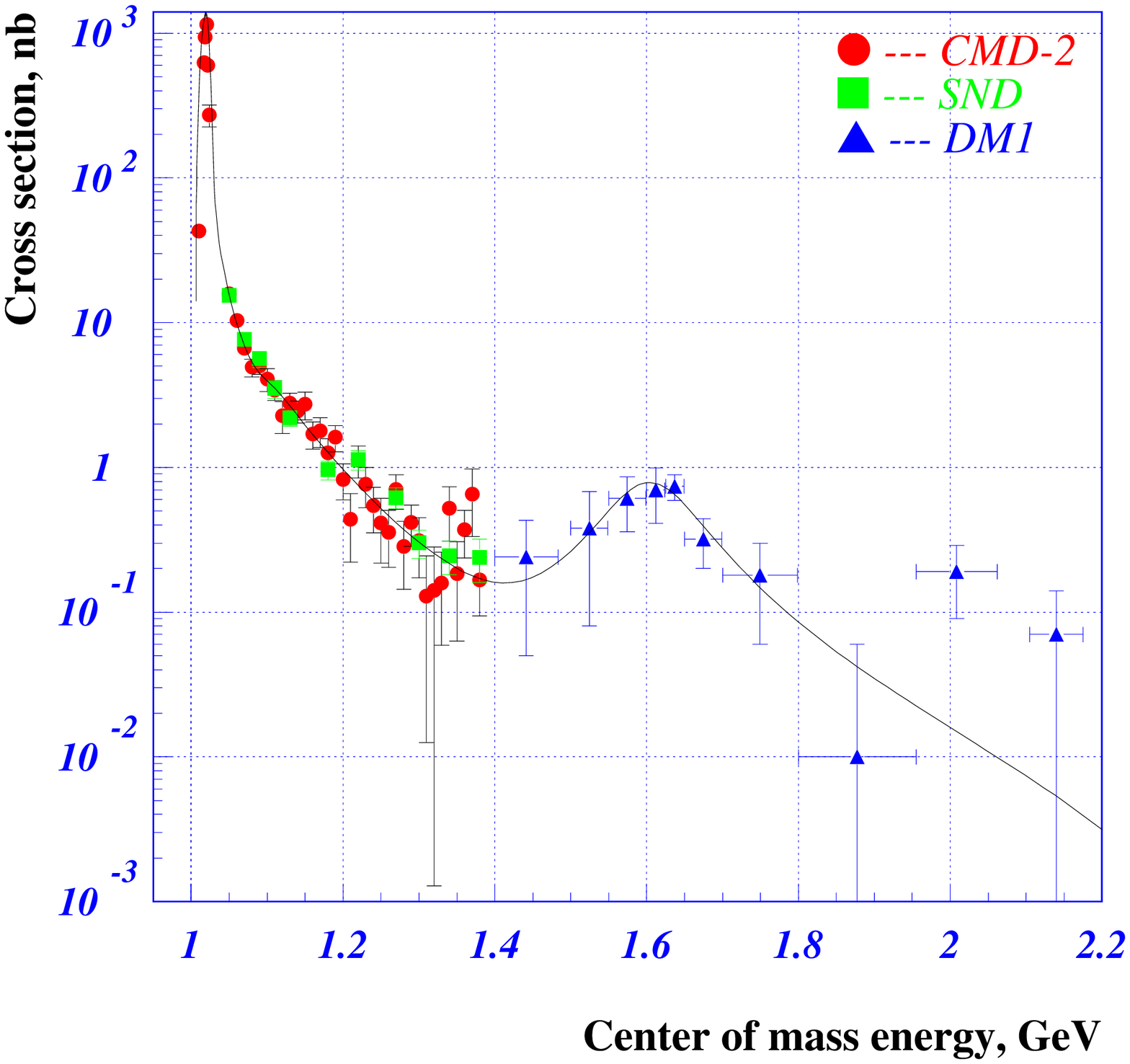} 
\caption
{The $e^+e^-\to K_SK_L$ cross section.} 
\label{kskl} 
\end{figure}

The preliminary SND results on 
the cross section 
$e^+e^-\to K_S K_L$
\cite{SNDphi,SNDrecent}
together with the DM1 data \cite{DM1KsKl} can be successfully fitted if 
the contributions of the $\rho$, $\omega$, $\phi$, and $\phi(1680)$
resonances are taken into account (Fig.\ref{kskl}).
The data on the reaction $e^+e^-\to K^+K^-$ \cite{OLYAK+K-, DM2K+K-}
are in agreement with a such model.

\subsection{$e^+e^-\to KK\pi$ cross sections}
 
The PWA analysis of the $e^+e^-\to KK\pi$ reaction
have been performed by DM2 \cite{DM2KKpi}.
It was shown that isoscalar process
$\phi(1680)\to K^*K\to K_SK^{\pm}\pi^0$ dominates. The cross
section  $e^+e^-\to K^+K^-\pi^0$ is small. The $1.45~GeV$ vector state 
observed in the hadron production \cite{phipi0}
is not confirmed in the 
$e^+e^-$ production at VEPP-2M \cite{SNDrecent,NDphipi0}. 

\section{THE LIGHT VECTOR MESON SPECTRUM} 

\begin{table}
\caption
{The classification of vector mesons by PDG'00.}
\begin{tabular}{|c|c|c|c|} 
\hline
$N^{2S+1}L_J$      &  $(u\overline{u}-d\overline{d})/\sqrt{2}$ &
         $(u\overline{u}+d\overline{d})/\sqrt{2}$ & $s\overline{s}$ \\ 
\hline
$1^3S_1$  & $\rho(770)$ & $\omega(782)$ & $\phi(1020)$ \\
\hline
$2^3S_1$  & $\rho(1450)$ & $\omega(1420)$ & $\phi(1680)$ \\
\hline
$1^3D_1$  & $\rho(1700)$ & $\omega(1650)$ & - \\
\hline
$3^3S_1$  & $\rho(2150)$ &  &  \\
\hline
\end{tabular} 
\label{t1}
\end{table}

\begin{table*}
\caption
{The level of experimental significance of the
vector mesons in selected reactions: + - well established states, 
* - not well established states.}
\begin{center}
\begin{tabular}{|l|c|c|c|c|c|c|c|c|c|} 
\hline
 &  $\rho(1250)$ & $\omega(1200)$  
 &  $\rho(1450)$ & $\omega(1420)$ & 
    $\rho(1700)$ & $\omega(1650)$ & $\phi(1680)$ & $\rho(2150)$ \\
\hline
$e^+e^-\to\pi^+\pi^-$      &* & & * &   &+&   &   & *  \\
$e^+e^-\to\pi^+\pi^-\pi^0$ &  &+ &   & * &   & * & * &    \\
$e^+e^-\to 4\pi$           &  & &+&   &+&   &   &   \\
$e^+e^-\to\omega\pi^0$     & & & * &   &+&   &   & \\
$e^+e^-\to\omega\pi^+\pi^-$& &* &   & * &   &+& * &  \\
$e^+e^-\to\eta\pi^+\pi^-$  & & &+&   & * &   &   &   \\
$e^+e^-\to\eta\gamma$      & & & * &   &   &   &   &   \\
$e^+e^-\to K_SK_L$         & & &   &   & * &   &+&   \\
$e^+e^-\to K^+K^-$         & & &   &   & * & * & * & + \\
$e^+e^-\to K^*K$           & & &   &   & * &   &+&   \\
$e^+e^-\to K^+K^-\pi^0$    & & &   &   &   &   & * &   \\
$\pi^-p\to \omega\pi^0 n$  & & &   &   &   &   &   & +  \\
$\pi^-p\to \phi\pi^0 n$    & & &   &   & * &   &   &   \\
$p\overline{p},\overline{p}n\to hadrons$    &*& & *  &   & + &   &   &   \\
$\gamma p\to hadrons$       &*& &   &   & + &   &   &   \\
\hline
\end{tabular} 
\end{center}
\label{t2}
\end{table*}

The classification (Table \ref{t1}) of the light vector 
mesons proposed 
by PDG \cite{PDG00} cannot be accepted without a serious 
discussion. Some of resonances included in the table are
no well established. In the contrary, the data on $\rho(1250)$ and 
$\omega(1200)$ are ignored. The difficulty of the existing
data analysis connect with the low statistical accuracy of the
data above $1.4~GeV$. Moreover the  model uncertainty of resonances mass
and width
may exceed $200~MeV$ \cite{Achasov}. The quality
of the experimental data is demonstrated in the Table \ref{t2} and
the following conclusions can be done after review
of the current data: 
\begin{itemize}
\item  
$\rho(1250)$ is ignored by PDG but as pointed out by D.~Peaslee
\cite{Peaslee} there are several old and new experiments 
(OMEGA \cite{OMEGA1}, LASS \cite{LASSpi2}, OBELIX \cite{OBELIX1,OBELIX2}) 
in which some evidences for the $\rho(1250)$ were obtained.
\item  
 $\omega(1200)$ is  identified by $\pi^+\pi^-\pi^0$
cross section \cite{SNDpi3}.
\item  
 $\rho(1450)$ is identified by $\pi^+\pi^-\pi^+\pi^-$
and $\eta\pi^+\pi^-$ production in $e^+e^-$ and in $p\overline{p}$
experiments.
\item  
 $\omega(1420)$ has no solid ground.
\item  
 $\rho(1700)$ is seen in the $e^+e^-$ production
in $\pi^+\pi^-$,
$\omega \pi^0$, and $\pi^+\pi^-\pi^0\pi^0$ final states.
It is  identified in the gamma production \cite{PDG00}
and in the $p\overline{p}$ production \cite{CB1, CB2}.  
\item  
 $\omega(1650)$ is identified by $\omega\pi^+\pi^-$
cross section \cite{DM2pi3}.
\item  
 $\phi(1680)$ is identified by the $K^*K$ 
cross section \cite{DM2KKpi}.
\item  
 $\rho(2150)$ is identified 
in the hadron production of $\omega\pi^0$ by GAMS \cite{GAMS1}.
\end{itemize}

\section{PROSPECTS FOR PEP-N}

There are a set of questions which must be answered to
clear the situation with excited states of the light vector mesons:  
\begin{itemize}
\item  
 Do $\rho(1250)$ exist? What is the nature of this object?
It is $2^3S_1$ $q\overline{q}$ state
or it is lowest 4-quark vector state?
\item  
 Do $\omega(1200)$ is $2^3S_1$ $q\overline{q}$ state
or it is lowest 4-quark vector state?
\item  
 Do $\omega(1420)$ exist?
\item  
 $\rho(1700)$, $\omega(1650)$, and $\phi(1680)$ have
practically the same mass. They have to have common decay channels,
so its real inputs are hidden in cross section shapes
because of the interference. 
Are there three resonances $\rho(1700)$, 
$\omega(1650)$, and $\phi(1680)$ 
or there are only two?
\item  
Do other light quarks states exist?
\end{itemize}
The adequate $e^+e^-$ collider for a such study is PEP-N.
The experiment at PEP-N is able to provide
a good efficiency and particle identification 
for hadron and radiative transitions between
different states in the energy region $1 - 3~GeV$. 
The other methods using existing facilities are not able to solve
all problems 
of the spectroscopy of the light vector mesons  
because of following problems:
\begin{itemize}
\item  
Below $3~GeV$ the luminosity of existing $e^+e^-$ colliders
fall down. The designed maximum energy of the VEPP-2000 \cite{Koop} is
$2~GeV$.
\item  
Using the hadronic $\tau$ decays is not possible to 
establish the spectrum of vector mesons above $1.3~GeV$
because of kinematics.
\item  
The Initial State Radiation method \cite{ISR, Solodov}
is a very effective method to demonstrate the cross section shape
and to tag the most interesting phenomena, but the precision of
this method is not known and
some theoretical and experimental limits can be foreseen. 
\item  
The previous experience shows us that experiments for  the hadron production,
 $\gamma$-production,
and $p\overline{p}$ production  cannot substitute 
precise $e^+e^-$ experiments for the vector meson spectroscopy.
\end{itemize}

\section{CONCLUSIONS} 

\begin{itemize}
\item  
The knowledge of the vector meson spectroscopy is incomplete.
\item  
The heavy quarkonium spectra are known much better
than the spectrum of the light vector mesons.
\item  
It is required 
to measure a complete set of hadron production
cross sections in the energy region $1 - 3~GeV$ with 
the integrated luminosity about $200~pb^{-1}$.
\item  
This luminosity investment will provide an opportunity
to study as traditional and exotic states,
hadronic and radiative transitions.
\item  
The two new $e^+e^-$ projects VEPP-2000 \cite{Koop} 
at Novosibirsk
and PEP-N at SLAC are intend to bring a light
on the light vector meson spectroscopy.
This two projects are complimentary in many aspects,
so the realisation of both is very well required.
\end{itemize}





\begin{thebibliography}{99} 
 
\bibitem{Sidorov} 
V.L.~Auslander et al., Yad. Fiz. 9: 114-119, 1969.

\bibitem{PDG00}
D.E.~Groom et al. (Particle Data Group), 
Eur. Phys. Jour. C15: 1, 2000. 

\bibitem{SNDpi0pi0g0}
M.N.~Achasov et al.,
Phys. Lett. B440: 442-448, 1998. 
e-Print Archive: hep-ex/9807016.

\bibitem{CMDpi0pi0g}
CMD-2 Collaboration (R.R.~Akhmetshin et al.), 
Phys. Lett. B462: 380, 1999.
e-Print Archive: hep-ex/9907006.

\bibitem{SNDpi0etag}
M.N.~Achasov et al.,
Phys. Lett. B479: 53-58, 2000. 
e-Print Archive: hep-ex/0003031.

\bibitem{SNDpi0pi0g}
M.N.~Achasov et al.,
Phys. Lett. B485: 349-356, 2000. 
e-Print Archive: hep-ex/0005017.  

\bibitem{Zallo} 
A.~Zallo, Talk at this workshop. 

\bibitem{OLYAK+K-}
P.M.~Ivanov et al.,  
Phys. Lett. B107: 297, 1981. 

\bibitem{OLYApi2}
L.M.~Barkov et al.,
Nucl. Phys. B256: 365-384, 1985.  

\bibitem{ND}
S.I.~Dolinsky et al., 
Phys. Rept. 202: 99-170, 1991. 

\bibitem{CMD2}
E.V.~Anashkin et. al., ICFA Instr. Bulletin 
5: 18, 1988.

\bibitem{SND}
M.N.~Achasov et al.,  
Nucl. Instrum. Meth. A449: 125-139, 2000. 
e-Print Archive: hep-ex/9909015.

\bibitem{SNDphi}
M.N. Achasov et al.,  
Phys. Rev. D63: 072002, 2001. 
e-Print Archive: hep-ex/0009036.

\bibitem{GS}
G.J.~Gounaris and J.J.~Sakurai,
Phys. Rev. Lett. 21: 244, 1968.

\bibitem{HLS}
M.~Bando et al.,
Phys. Rev. Lett. 54: 1215, 1985.

\bibitem{fresh}
N.N.~Achasov et al., 
Sov. J. Nucl. Phys. 54: 664-671, 1991;
Int. J. Mod. Phys. A7: 3187-3202, 1992. 

\bibitem{BEN1}
M. Benayoun, S.I. Eidelman, V.N. Ivanchenko, 
Z. Phys. C72: 221-230, 1996. 


\bibitem{Achasov}
N.N.~Achasov and A.A.~Kozhevnikov, Phys. Rev. D62: 117503, 2000.

\bibitem{CMDpi2}
CMD-2 Collaboration (R.R.~Akhmetshin et al.). BUDKERINP-99-10, Apr 1999. 52pp. 
e-Print Archive: hep-ex/9904027. 

\bibitem{Logoshenko} 
I.B.~Logoshenko, Talk at this workshop. 

\bibitem{DM2pi2}
D.~Bisello et al.,  
Phys. Lett. B220: 321, 1989. 

\bibitem{BEN2}
M.~Benayoun et al.,
Eur. Phys. J. C2: 269, 1998. 

\bibitem{DM2pi3}
A.~Antonelli et al.,  
Z. Phys. C56: 15-19, 1992. 

\bibitem{SNDpi3}
M.N.~Achasov et al.,  
Phys. Lett. B462: 365-370, 1999. 
e-Print Archive: hep-ex/9910001.

\bibitem{CMDa1pi}
CMD2 Collaboration (R.R.~Akhmetshin et al.), Phys. Lett. B466: 392-402, 1999. 
e-Print Archive: hep-ex/9904024.


\bibitem{SNDpi4}
M.N. Achasov et al.,  
Preprint IYaF  2001-34, Novosibirsk, 2001.

\bibitem{CMDpi5}
CMD-2 Collaboration (R.R.~Akhmetshin et al.),
Phys. Lett. B489: 125-130, 2000. 
e-Print Archive: hep-ex/0009013.

\bibitem{SNDomegapi0}
M.N.~Achasov et al.,  
Phys. Lett. B486: 29-34, 2000. 
e-Print Archive: hep-ex/0005032. 

\bibitem{DM2omegapi0}
D.~Bisello et al.,  
Nucl. Phys. Proc. Suppl. 21: 111, 1991. 

\bibitem{CLEOomegapi0}
K.W.~Edwards et al.,  
Phys. Rev. D61: 072003, 2000. 
e-Print Archive: hep-ex/9908024. 

\bibitem{CMDetag}
CMD-2 Collaboration (R.R.~Akhmetshin et al.), 
Submitted to Phys. Lett. B.
e-Print Archive: hep-ex/0103043. 

\bibitem{DM2etapipi}
A.~Antonelli et al.,  
Phys. Lett. B212: 133, 1988. 

\bibitem{SNDrecent}
M.N.~Achasov et al.,  
Nucl. Phys. A675: 391c-397c, 2000. 
e-Print Archive: hep-ex/9910057. 


\bibitem{DM1KsKl}
J.~Buon et al.,  
Phys. Lett. B118: 221, 1982. 


\bibitem{DM2K+K-}
D.~Bisello et al.,  
Z. Phys. C39: 13-19, 1988. 

\bibitem{DM2KKpi}
D.~Bisello et al.,  
Z. Phys. C52: 227-230, 1988. 

\bibitem{phipi0}
S.I.~Bityukov et al.,  
Phys. Lett. B188: 383, 1987. 

\bibitem{NDphipi0}
V.M. Aulchenko et al., 
JETP Lett. 45: 145-147, 1987; 
Pisma Zh. Eksp. Teor. Fiz. 45: 118-120, 1987. 

\bibitem{Peaslee} 
D.C.~Peaslee, Private communication. 

\bibitem{OMEGA1} 
M.~Atkinson et al., 
Nucl. Phys. B245: 189, 1984.  

\bibitem{LASSpi2}
D.~Aston et al., SLAC-PUB-5722, Dec 1991. 
In *College Park 1991, Proceedings, Hadron '91* 75-78. 

\bibitem{OBELIX1}
OBELIX Collaboration (A. Bertin et al.), 
Phys. Lett. B408: 476, 1997. 

\bibitem{OBELIX2}
OBELIX Collaboration (A. Bertin et al.), 
Phys. Lett. B414: 220-228, 1997. 

\bibitem{CB1}
A.~Abele, 
Phys. Lett. B391: 191, 1997. 

\bibitem{CB2}
A.~Abele, 
Phys. Lett. B468: 178, 1999. 

\bibitem{GAMS1}
D.M.~Alde et al., 
Z. Phys. C66: 379, 1995. 

\bibitem{Koop} 
I.A.~Koop, Talk at this workshop. 

\bibitem{ISR} 
M.~Benayoun et al.,
Mod. Phys. Lett. A14: 2605-2614, 1999. 

\bibitem{Solodov} 
E.P.~Solodov, Talk at this workshop. 

 
\end{thebibliography}
\end{document}